\documentclass[prb,twocolumn,superscriptaddress,preprintnumbers,amsmath,amssymb]{revtex4}
\usepackage{color}
\usepackage{graphicx}
\usepackage{amsmath}
\usepackage{amssymb}
\usepackage{epsfig}
\usepackage{wasysym}
\usepackage{bbm}
\usepackage{multirow}
\renewcommand{\narrowtext}{\begin{multicols}{2} \global\columnwidth20.5pc}
\renewcommand{\v}[1]{{\bf #1}}

\newcommand{\s}{{\sigma}}
\newcommand{\bz}{\bar{z}}

\def\be{\begin{eqnarray}}
\def\ee{\end{eqnarray}}

\newcommand{\Eq}[1]{Eq.~(\ref{#1})}

\renewcommand{\>}{\rangle}

\newcommand{\p}{\partial}

\newcommand{\ra}{\rightarrow}

\newcommand{\e}{\epsilon}
\newcommand{\Fig}[1]{Fig.~(\ref{#1})}

\begin{document}

\title{The surface states of topological insulators - Dirac fermion in curved two dimensional spaces}


\author{Dung-Hai Lee}
\affiliation{
Department of Physics,University of California at Berkeley,
Berkeley, CA 94720, USA}
\affiliation{Materials Sciences Division,
Lawrence Berkeley National Laboratory, Berkeley, CA 94720, USA}

\begin{abstract}
The surface of a topological insulator is a closed two dimensional manifold. The surface states are described by the Dirac Hamiltonian in curved two dimensional spaces. For a slab-like sample with a magnetic field perpendicular to its top and bottom surfaces, there are chiral states delocalized on the four side faces. These ``chiral sheets'' carry both charge and spin currents. In strong magnetic fields, the quantized charge Hall effect ($\s_{xy}=(2n+1)e^2/h$) will coexist with spin Hall effect.

\end{abstract}

\date{\today}
\maketitle

\begin{figure}[tbp]
\begin{center}
\includegraphics[angle=0,scale=0.33]
{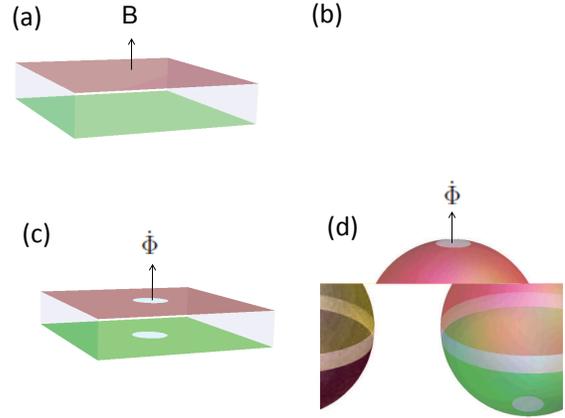}\caption{(color on-line)  The surface of a slab (a) and its topological equivalent - the surface of a sphere (b). The red and green colors indicate the signs of the magnetic flux being opposite.  The white region has no magnetic flux piercing. (c),(d) the current induced by flux ramping. In part (d) the dashed circles are drawn for keeping track of the charge that flows pass it.  \label{dipolar}}
\end{center}
\end{figure}

The theoretical prediction that 3D topological insulators necessarily possess massless Dirac fermion surface states\cite{topo}, and the fact that they are observed in real materials\cite{ARPES} have stimulated lots of interest recently. In considering the surface electronic structure of such materials  it is important to remember that the surface of a bulk sample is a closed two dimensional manifold. For example, the surface of a slab has the topology of a sphere (\Fig{dipolar}{a}). If we apply an upward magnetic field the net flux passing through the surface is zero. The situation is shown in \Fig{dipolar}(a) where the magnetic field at the top surface points along the surface normal (red), while that at the bottom  points opposite to the surface normal (green). Since there is no magnetic flux piercing the four side faces, they are left white. \Fig{dipolar}(b) is topologically equivalent to \Fig{dipolar}(a), and the presence of azimuthal symmetry makes it a convenient geometry for theoretical studies.
Because the net magnetic flux is zero there is no topologically protected zero energy states\cite{as}. This raises  questions about the integrity of the Landau levels and the existence of the quantized Hall effect.

What is the electronic structure of a topologically insulator with the shape of \Fig{dipolar}(a)? Since the  Landau-level states on the top and bottom surfaces are connected by delocalized states on the side faces, it is not clear what the answer should be. It is well known that in an infinite plane each Dirac point contributes  $ (n+1/2)e^2/h$ to the Hall conductance. What should the Hall conductance of a slab be? In addition, because a closed surface has no edges, the validity of the conventional edge state picture for integer quantum Hall effect\cite{halperin} is questionable even if the quantized Hall effect does exist. Finally if, for $B=0$, there is one Dirac point on the surface of a semi-infinite sample, how many Dirac points should the surface of a slab have? These questions motivate us to look at the surface electronic structure of a finite topological insulator more closely.

In the following we shall focus on the simplest case where the surface Dirac point is derived from of the bulk
electronic structure near $\vec{k}=(0,0,0)$ (interestingly this is the situation of Bi$_2$Se$_3$\cite{bi2se3}).
We begin by asking the following question. ``If
$H=\sum_{k=1,2} \s_k(-i\p_k-A_k)$ describes the surface electronic structure of an semi-infinite topological insulator what is the surface Hamiltonian for a finite slab-like sample?'' Here $k=1,2$ label the two orthogonal coordinate frames in the surface;  $A_k$ is the component of the electromagnetic vector potential; $\vec{\s}$ are the Pauli matrices, and  $(\hbar/2)\hat{n}\times\vec{\s}$, where $\hat{n}$ is the surface normal, is the in-plane physical spin operator.
For simplicity, we have set the Fermi velocity to unity. A clarification is in order. For a real material, crystalline anisotropy usually results in anisotropy in the low energy effective theory. For example the bulk effective Hamiltonian given in Ref.\cite{stanford} clearly exhibits a different Fermi velocity in the $z$ direction from that in the $x,y$ directions. However, this difference can be removed by a rescaling of the $z$ coordinate.

We note that any surface with the topology of a sphere has total Gauss curvature $4\pi$ hence is necessarily curved. It can be shown,  by, e.g., projecting the bulk effective Hamiltonian in Ref.\cite{stanford} onto surfaces of different orientations, that the ``spin'' $\vec{\s}$ of the surface states is always locked within the surface plane. Due to this spin-space locking, the appropriate Hamiltonian  is the Dirac Hamiltonian in curved spaces.

The curvature of the space enters through a non-abelian spinor connection\cite{witten}. (Note in our case only the space is curved not the time.) In the following we shall choose the most convenient coordinate system to write down such a Hamiltonain.  It has been proven\cite{chern} that for any Riemann surface it is possible to choose a coordinate system such that the metric is conformally related to the Euclidean metric everywhere, i.e.,
\be
ds^2=e^{-2\beta(\v x)}\left[(dx^1)^2+(dx^2)^2\right].\label{metric}\ee Such a coordinate system is called ``isothermal''.
In terms of isothermal coordinates the Dirac Hamiltonian read \be
 H=e^{\beta (\v x)}\sum_{k=1,2}\s_k(\v x)\left(-i \partial_k+ i \Gamma_k(\v x)-iA_k(\v x)\right).\label{cd}\ee
 Here  $\s_k(\v x)\equiv\hat{e}_k(\v x)\cdot\vec{\s}$ with $\hat{e}_{1,2}(\v x)$ being the local orthonormal coordinate frame so that $\hat{e}_3(\v x)=\hat{e}_1(\v x)\times\hat{e}_2(\v x)=\hat{n}$ is the unit surface normal. The spinor connection $\Gamma_k(\v x)$ is given by
 \be
\Gamma_k(\v x)={1\over i}\left[\sum_j\e_{kj}\p_j\beta(\v x)\hat{e}_3(\v x)+\frac{1}{2}\hat{e}_3(\v x)\times\partial_k \hat{e}_3(\v x)\right]\cdot\vec{\s}\nonumber\ee
(Note that due to the choice of isothermal coordinates we have used the usual Euclidean space notion which does not distinguish the covariant and contravariant indices.)
  If we define $|z_\pm(\v x)\>$ as the eigenvectors of $S_3(\v x)$, i.e., $S_3(\v x)|z_\pm(\v x)\>=\pm |z_\pm(\v x)\>$, and sandwich the Hamiltonian in
 \Eq{cd} between them, the following $2\times 2$ matrix operator is obtained\cite{pnueli}
 \be
 H=-2i e^{\beta}\left(
\begin{array}{cc}
 0 & \left( \partial _z-i A_z\right)+\frac{1}{2}\partial _z \beta \\
 \left( \partial _{\bar{z}}-i A_{\bar{z}}\right)+\frac{1}{2}\partial _{\bar{z}} \beta & 0
\end{array}
\right)\label{h0}
 \ee
In the above equation  $ z=x^1+ix^2,\bar{z}=x^1-ix^2,\p_z=1/2(\p_1-i\p_2), \p_{\bar{z}}=1/2(\p_1+i\p_2)$, and $A_z=1/2(A_1-iA_2), A_{\bar{z}}=1/2(A_1+iA_2)$, respectively.
In the following we shall solve the eigenvalue problem
with the magnetic field and the configuration space shown in \Fig{dipolar}(b) and the Hamiltonian given by \Eq{h0}.

First, we choose the isothermal coordinate system $(u,\phi)$ where each point $\v r\in S^2$ is parameterized as
$\v r(u,\phi)=R\left(\sin\alpha(u)\cos\phi,\sin\alpha(u)\sin\phi,\cos\alpha(u)\right).$
Here $R$ is the radius of the sphere, $-\infty<u<\infty$, $0\le\phi<2\pi$ and $\alpha(u)=2\tan^{-1}(e^u)$. As $u\ra\pm\infty$ the point $\v r(u,\phi)$ approaches the south and north pole, respectively. Under this coordinate system  \be e^{-2\beta(z,\bar{z})}=R^2{\rm sech}^2(u),\label{beta}\ee
where $z=u+i\phi$,
 and the Gauss curvature is given by $G(z,\bz)=4 e^{2\beta}\partial _z \partial _{\bar{z}} \beta$.
 In the same complex notation the magnetic field is given by $B=-2i e^{2\beta}(\p_zA_{\bar{z}}-\p_{\bar{z}}A_z)$.
If the magnetic field is uniform (hence requires a magnetic monopole to generate it) the field strength is related to $G$ by
$B=n_\phi G/2$ where $n_\phi$ is the number of flux quanta piercing through the sphere.
If we pick the gauge $\p_1A_1+\p_2A_2=0$, a $\Lambda(z,\bz)$ can be found so that 
$A_z=-i\p_z\Lambda,~A_{\bar{z}}=i\p_{\bar{z}}\Lambda$ and $B=4 e^{2\beta}\partial _z \partial _{\bar{z}} \Lambda$. Compare this equation with $G=4 e^{2\beta}\partial _z \partial _{\bar{z}} \beta$ and $B=n_\phi G/2$ suggests that we can set
\be
\Lambda(z,\bar{z})={n_\phi\over 2}\beta(z,\bar{z}).\label{lam}\ee By substituting $A_z=-i\p_z\Lambda,A_{\bar{z}}=i\p_{\bar{z}}\Lambda$ and \Eq{lam} into \Eq{h0} we obtain, \be
H=-2i e^{\beta}\left(
\begin{array}{cc}
 0 & \partial _z+(\frac{1}{2}-n_\phi)\partial _z \beta \\
 \partial_{\bar{z}}+(\frac{1}{2}+n_\phi)\partial _{\bar{z}}\beta & 0
\end{array}
\right)\label{h}
 \ee
\begin{figure}[tbp]
\begin{center}
\includegraphics[angle=0,scale=0.33]
{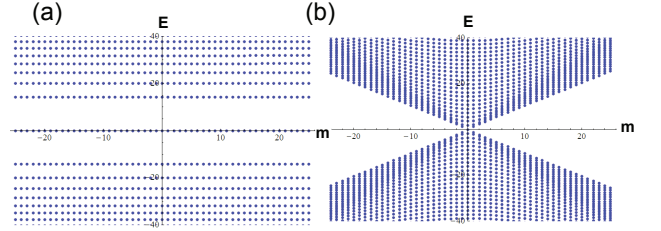}\vspace{-0.8in}\caption{(color on-line) The eigen spectra of \Eq{eigen} under constant magnetic field - (a), and zero magnetic field - (b).  \label{landau}}
\end{center}
\end{figure}

Because of \Eq{beta} the Hamiltonian in \Eq{h} is $\phi$-translation invariant, the eigenfunctions are of the form
$\psi^T=e^{i m \phi }(u_m,v_m),$
where $m$, the azimuthal quantum number, is an integer. The equation satisfied by $u_m$ and $v_m$ is
\begin{widetext}
\be e^{\beta(u)}\left[-is_x\left(\partial_u+\frac{1}{2}\beta^\prime(u)\right)+s_y\left(m-{n_\phi\over 2}\beta^\prime(u)\right)\right]\left(
\begin{array}{c}
 u_m \\
 v_m
\end{array}
\right)=E\left(
\begin{array}{c}
 u_m \\
 v_m
\end{array}
\right).
\label{eigen}
\ee
\end{widetext}
In the absence of magnetic field ($n_\phi=0$) the eigenspectrum is shown as \Fig{landau}(b). At $m=0$ there is {\it one} pair of eigenvalues  with $E=0$, thus there is only one Dirac point! For $n_\phi>>1$ the eigen spectrum of \Eq{eigen} consists of energy levels which are independent of $m$: $E_n=\pm\sqrt{n(n_\phi+n)/R^2}$ where $n\ge 0$ as shown in the inset of \Fig{landau}(a).
This is the relativistic Landau level on a sphere\cite{pnueli}.

\begin{figure}[tbp]
\begin{center}
\includegraphics[angle=0,scale=0.33]
{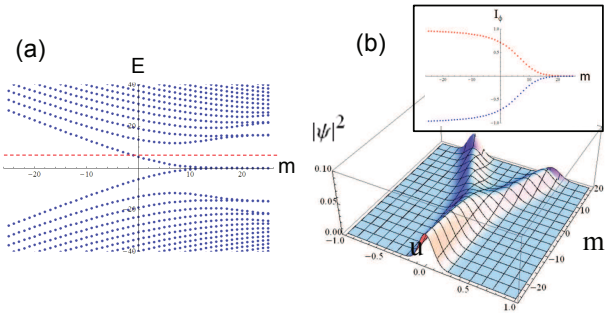}\vspace{-0.4in}\caption{(color on-line) (a) The eigen spectrum of \Eq{eigen} with magnetic field shown in \Fig{dipolar}(b). The red dashed line denotes the Fermi level. (b) $|u_m(u)|^2+|v_m(u)|^2$ as a function of $u$ and $m$ for the upper $n=0$ band (the band intersected by the Fermi level). Insert of part (b) the The current ($I_\phi$) expectation value associated with the (split) $n=0$ Landau band.  \label{results}}
\end{center}
\end{figure}

Next, we consider the magnetic field illustrated in \Fig{dipolar}(b). Here we choose
\be
\Lambda(u)={n_\phi\over 2}\left(\tanh{u\over w}\right)\beta(u),\label{lam1}\ee  where $w$ is proportional to the width of the zero-flux band in \Fig{dipolar}(a). For the results shown in \Fig{results} $w$ is set equal to the magnetic length.  \Fig{results}(a) is the eigen spectrum.  For large $m$ the Landau levels in \Fig{landau}(a) is recovered. However unlike
\Fig{landau}(a) where the energy levels are non-degenerate at each $m$, in \Fig{results}(a), as $m\ra\infty$, each energy level is two fold-degenerate. This degeneracy arises from the Landau states in  the two hemispheres. In the presence of the Zeeman term the Landau level will be split even as $m\ra\infty$. Usually that splitting is much smaller than the Landau gap hence will not affect any of the discussions below.

In the following let us  focus on the $n=0$ Landau level.
As $m$ decreases the degeneracy are split due to the mixing with the delocalized states around the equator. Since the overlap between the Landau states and the delocalized states vanishes exponentially as $m\ra\infty$ so does the  splitting. However, strictly speaking so long as $m<\infty$ the splitting is always non-zero.
This is consistent with the fact that there is no topologically protected zero modes.
Spatially the eigenfunctions are localized in the zero-flux band around the equator when $m$ is sufficiently small (or negative). At large $m$ the eigen functions are equal amplitude superpositions of two components localized in different hemispheres(\Fig{results}(b)). This is due to the mixing discussed above. In the inset of \Fig{results}(b) we show the expectation value of the current operator, $I_\phi\ra 0$ for large $m$ and develop positive/negative values for the upper/lower band as $m$ decreases.   If the Fermi level is marked by the dashed red line in \Fig{results}(a), it will intersect the chiral, current-carrying states. These are the ``edge'' states of our problem. Their dispersion are remarkably similar to the valley-split edge states in graphene\cite{ab}.
The current in the chiral state is distributed across the width of the zero-flux band.
Thus we have chiral sheets.

Now we switch to the existence of quantum Hall effect. Let us imagine inserting a thin solenoid as shown in \Fig{dipolar}(d) (or (c)) and slowly ramping its flux from $0$ to $\phi_0$. An adiabatic evolution, $m\ra m-1$, will be induced. At the end, an extra electron will appear above the Fermi level. At the mean time the charges reside in each large $m$ states march toward the zero-flux band around the equator. Since these states have $50\%$ weight in each hemisphere (\Fig{dipolar}(a)), each dashed circle in \Fig{dipolar}(d) will register an average $1/2$ of an electron
crossing it. This suggests that each hemisphere transfers $1/2$ of an electron to the equator. The fact that an azimuthal electric field induces a quantized charge accumulation in the perpendicular direction suggests the existence of quantized Hall effect. The fact $1/2+1/2=1$ electron is transferred implies  $\s_{xy}=e^2/h.$ If the Fermi level intersects the lower $n=0$ band, the direction of current will be reversed, hence $\s_{xy}=-e^2/h$.
Similar argument carried out for higher Landau levels gives $\s_{xy}=(2n+1)e^2/h$ where $n$ are integers. These are exactly half of the quantized Hall conductances observed in graphene\cite{yb}.
The above result is consistent with the statement that $\s_{xy}=\pm (n+1/2)e^2/h$ for the two hemispheres.(Here the sign of the Hall conductance is defined in reference to the surface normal so that $\vec{j}=\sigma_{xy}\hat{n}\times\vec{E}$).
The fact that the induced EMF circulates the surface normal in opposite senses in the two hemispheres, and the fact that the Hall conductances have opposite sign, imply the induced Hall current will flow toward (or away from) the equator in both hemispheres. Because $|\sigma_{xy}|=(n+1/2)e^2/h$, the total charge that flows to/from  the equator is $2\times (n+1/2)=2n+1$ electrons.
\begin{figure}[tbp]
\begin{center}
\hspace{-.5 in}
\includegraphics[angle=0,scale=0.45]
{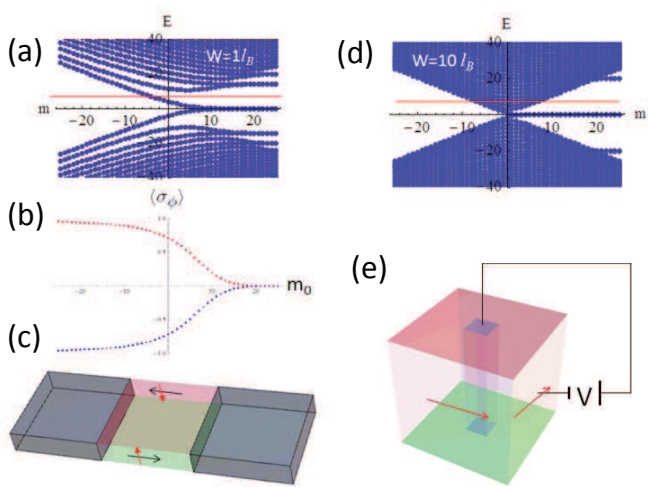}\caption{(color on-line) (a) The spectrum of the configuration \Fig{dipolar}(b) for $w$ (the width of the zero flux band) equal to the magnetic length. (b) The $\s$ polarization of the chiral sheet states as a function of $m_0$. (c) A two terminal device. Here the arrows indicate the direction of charge as well as spin currents.(d) The spectrum of the configuration \Fig{dipolar}(b) for $w$ equal to ten magnetic length. (e) The set up for measuring the quantized Hall effect in the limit thick slab.\label{spin}}
\end{center}
\end{figure}

The chiral sheet current discussed above
also carries spin polarization. The effective field theory describing the chiral state is given by
$
H_{chiral}=
-i\int d\phi \chi^\dagger(\phi)\p_\phi\chi(\phi).$ 
In the slab geometry $\phi$ is the coordinate around the four side faces.
The spinless fermion operator  $\chi(\phi)$ is related to the original spinful fermion operator $\psi_\s(u,\phi)$ via
$
\chi(\phi)=\int du \sum_\s \xi_{0,\s}(u,\phi)\psi_\s(u,\phi).$ In the above equation  $\xi_{0,\s}(u,\phi)=u_{m_0}(u)\langle\s|z_+(u,\phi)\rangle+
v_{m_0}(u)\langle\s|z_-(u,\phi)\rangle$, where $(u_{m_0}(u),v_{m_0}(u))$ is the solution of \Eq{eigen} for the upper upper $n=0$ band at $m=m_0$, and $|z_\pm(u,\phi)\rangle$ are the eigen spinors of $\s_3(u,\phi)=\hat{e}_3(u,\phi)\cdot\vec{\s}$. In \Fig{spin}(b) we plot the expectation value
$\langle \s_\phi\rangle=\int du~\hat{e}_2(u,\phi)\cdot\langle \xi_0(u,\phi)|\vec{\s}|\xi_0(u,\phi)\rangle
=i\int du\bar{v}_{m_0}(u)u_{m_0}(u)+ c.c$ as a function of $m_0$. Clearly, these chiral states also carries a $\vec{\s}$ polarization parallel to its current direction. However, because the physical spin is $(\hbar/2)\hat{n}\times\vec{\s}$, the actual spin polarization will be perpendicular to the current direction.  This is illustrated in \Fig{spin}(c), where the direction of the average spin polarization and the charge current on the side surfaces are indicted by the red and black arrows, respectively. Thus in the case of thin slab (i.e. thickness $\sim$ magnetic length) a two-terminal measurement (\Fig{spin}(c)) will exhibit both quantized charge conductance and spin Hall effect. 

So far we have set the width of the zero-flux band to the magnetic length. As the width increases, the edge states quickly merge into a continuum, as shown in \Fig{spin}(d) when $w=10 l_B$. However so long as the Fermi energy lies between the $n$ and $n+1$ Landau level away from the zero-flux band (see the red line in \Fig{spin}(d)) the previous flux ramping argument still implies the flow of $2n+1$ electrons into/from the zero-flux band upon the adiabatic evolution $m\ra m+1$. This suggests that in midst of the continuum states, there are a net $2n+1$ chiral current-carrying edge channels. Due to the anti-localization effect\cite{ryu}, both the chiral and the continuum states will be delocalized. In this case the two-terminal conductance will no-longer be quantized. In order to observe the quantized Hall effect one has to set up a ``Corbino'' geometry as shown in \Fig{spin}(e). In this case the circulating current on the side surface (indicated by the red arrow) will be $(2n+1)e^2/h$ times the voltage between the electrode and the outer surface of the sample.

In summary, we solve the curved space Dirac equation to study the surface electronic structure of topological insulators. For a slab-like sample with magnetic field perpendicular to its top and bottom surfaces, there are chiral states delocalized on the four side faces. These ``chiral sheets'' carry charge and spin currents. We discuss the
experimental to measure the quantized charge Hall effect ($\s_{xy}=(2n+1)e^2/h$) in the limit of thin and thick slabs. Due to the spin-orbit interaction the charge current carries spin polarization. In thick slabs, due to the anti-localization effect both the chiral and non-chiral electronic states will be extended on the side surfaces. These states will exhibit  unconventional Aharonov-Bohm effect in weaker field, when phase coherence is achieved.

{\bf Acknowledgments:~} This research was stimulated by the discussion with Geoffrey Lee on the relation between the Riemann-Roch theorem and  the Dirac equation. 
DHL was supported by DOE grant number DE-AC02-05CH11231.

\end{document}